\documentclass[twocolumn,10pt]{article}
\usepackage[utf8]{inputenc}
\usepackage[T1]{fontenc}
\usepackage[american]{babel}
\usepackage{amsmath,amssymb}
\usepackage{graphicx}
\usepackage{float}
\usepackage{array}
\usepackage{booktabs}
\usepackage{multirow}
\usepackage{url}
\usepackage{natbib}
\usepackage[margin=1in]{geometry}

\usepackage{tabularx}
\usepackage{ltxtable}
\usepackage{longtable}

\newcolumntype{L}[1]{>{\raggedright\arraybackslash}p{#1}}
\newcolumntype{C}[1]{>{\centering\arraybackslash}p{#1}}
\newcolumntype{R}[1]{>{\raggedleft\arraybackslash}p{#1}}

\title{CIA+TA Risk Assessment for AI Reasoning Vulnerabilities}

\author{
Yuksel AYDIN\\
Independent Researcher\\
\texttt{}
}

\date{}

\begin{document}

\maketitle

\begin{abstract}
As AI systems increasingly influence critical decisions, they face threats that exploit reasoning mechanisms rather than technical infrastructure. We present a framework for cognitive cybersecurity, a systematic protection of AI reasoning processes from adversarial manipulation. Our contributions are threefold. First, we establish cognitive cybersecurity as a discipline complementing traditional cybersecurity and AI safety, addressing vulnerabilities where legitimate inputs corrupt reasoning while evading conventional controls. Second, we introduce the CIA+TA, extending traditional Confidentiality, Integrity, and Availability triad with Trust (epistemic validation) and Autonomy (human agency preservation), requirements unique to systems generating knowledge claims and mediating decisions. Third, we present a quantitative risk assessment methodology with empirically-derived coefficients, enabling organizations to measure cognitive security risks. We map our framework to OWASP LLM Top 10 and MITRE ATLAS, facilitating operational integration. Validation through previously published studies (151 human participants; 12,180 AI trials) reveals strong architecture dependence: identical defenses produce effects ranging from 96\% reduction to 135\% amplification of vulnerabilities. This necessitates pre-deployment Cognitive Penetration Testing as a governance requirement for trustworthy AI deployment.
\end{abstract}

\section{Introduction}

The integration of large language models and reasoning-capable AI systems into critical organizational functions introduces attack surfaces that operate through fundamentally different mechanisms than traditional cybersecurity threats. When artificial intelligence systems acquire capabilities for contextual reasoning, multi-turn dialogue, and autonomous decision-making, they become vulnerable to adversarial manipulation through crafted information inputs that exploit systematic weaknesses in how these systems process and synthesize information. These cognitive vulnerabilities persist even in technically secure systems with properly configured access controls, encrypted communications, and validated software implementations.

Consider a financial AI system analyzing market data for investment decisions. Traditional security measures protect against unauthorized access, data tampering, and service disruption. However, an adversary can achieve their objectives by subtly manipulating the reasoning process itself, introducing biased information across multiple interactions that gradually shifts the system's analytical stance, embedding false contextual claims that the system incorporates as ground truth, or exploiting emotional framing that overrides logical analysis. The system continues operating normally by all technical measures while producing compromised outputs that reflect the attacker's intent rather than objective analysis. These vectors include indirect prompt injection via untrusted content and multi-turn prompt interference, now demonstrated in the wild and benchmarked systematically \cite{greshake2023indirect, yi2023bipia}. These vulnerabilities build upon foundational research in adversarial machine learning \cite{goodfellow2015explaining, carlini2017towards}, but extend beyond traditional adversarial examples to encompass sophisticated manipulation of reasoning processes \cite{zou2023universal}.

The severity of this risk becomes apparent through recent documented incidents. Recent research has demonstrated systematic vulnerabilities in language models, including sycophancy \cite{sharma2023towards}, hallucination \cite{xu2024hallucination}, and susceptibility to prompt injection attacks \cite{liu2023prompt, liu2024automatic}. Models sometimes generate fabricated citations under pressure; we observed architecture-dependent degradation in DOI validity and, in some mitigated settings, refusals to produce citations. Extended conversational interactions can produce stance drift across turns (quantified by per-turn slopes rather than SD/turn). Systems presented with conflicting objectives may yield inconsistent or incomplete outputs. These failures occur not through software bugs or misconfigurations but through exploitation of inherent characteristics in how AI systems reason about information.

Traditional cybersecurity frameworks prove insufficient for addressing these cognitive threats for three primary reasons. First, the attack vectors operate through legitimate channels using authorized inputs, bypassing conventional security controls entirely. Second, the exploitation targets reasoning patterns and information synthesis mechanisms rather than technical vulnerabilities that security tools are designed to detect. Third, the impacts manifest as degraded decision quality or manipulated outputs rather than system crashes or data breaches that trigger traditional incident response.

This paper signals cognitive cybersecurity as a complementary discipline requiring specialized frameworks for understanding, quantifying, and mitigating these reasoning-level vulnerabilities. We present three primary contributions:

\textbf{First}, we provide a formal definition of cognitive cybersecurity and identify seven fundamental vulnerability categories (CCS-7) that characterize the cognitive attack surface. This taxonomy, validated through extensive empirical testing, demonstrates that cognitive vulnerabilities exhibit strong architecture-dependent patterns that preclude universal defensive strategies.

\textbf{Second}, we introduce a quantitative risk assessment framework that enables organizations to measure and prioritize cognitive security risks based on empirical data rather than subjective estimates. The framework incorporates architecture-specific modifiers and mitigation effectiveness coefficients derived from 12,180 controlled experiments across seven distinct AI architectures.

\textbf{Third}, we present the CIA+TA security framework that extends traditional confidentiality, integrity, and availability principles with two additional components essential for cognitive systems: Trust (epistemic validation of knowledge claims) and Autonomy (preservation of human agency in AI-mediated decisions). This extended framework addresses security requirements unique to systems with artificial reasoning capabilities that generate knowledge claims and influence human decision-making.

Our empirical foundation draws from two studies. The first study validated proposed mitigations through human subject experiments with 151 participants, confirming that certain interventions successfully reduce human susceptibility to AI-mediated manipulation. The second examined cognitive vulnerabilities across trials for each of seven architectures (from 630 to 3,150 trials), revealing that identical defensive measures are mostly effective but can produce opposite effects depending on system design, with mitigation coefficients ranging from $\eta = 0.96$ (96\% reduction in vulnerability) to $\eta = -1.35$ (135\% increase).

The implications extend beyond technical security to encompass organizational resilience and strategic risk. As AI systems increasingly influence high-stakes decisions in healthcare, finance, critical infrastructure, and everyday life, the potential for adversarial manipulation through cognitive attack vectors represents a systemic risk that current security practices do not adequately address yet. Organizations deploying AI systems without cognitive security measures face exposure to sophisticated attacks that compromise decision-making integrity while evading all traditional security controls.

\section{Defining Cognitive Cybersecurity}

\subsection{Formal Definition and Scope}

Cognitive cybersecurity addresses systematic vulnerabilities in artificial reasoning processes that create exploitable attack vectors through information manipulation rather than system compromise. This discipline focuses on threats that target the mechanisms by which AI systems synthesize knowledge, maintain contextual understanding, and generate decisions. Unlike vulnerabilities in code or infrastructure, cognitive vulnerabilities arise from the fundamental characteristics of how these systems process information, making them inherent to the reasoning capability itself rather than its implementation. This definition extends beyond traditional AI safety concerns \cite{amodei2016concrete} and adversarial robustness \cite{goodfellow2015explaining} to address the unique challenges posed by reasoning-capable systems that influence human decision-making.

The scope encompasses AI systems that demonstrate three critical capabilities: autonomous reasoning across domains, contextual retention across interactions, and influence over consequential decisions. These systems become vulnerable precisely because they replicate aspects of human cognition, inheriting systematic biases and reasoning patterns that adversaries can exploit. The attack surface emerges at the intersection of system capability and information processing, where legitimate inputs can be crafted to corrupt reasoning while remaining within normal operational parameters.

\subsection{Positioning with Adjacent Disciplines}

Cognitive cybersecurity can best be understood as a complementary angle of analysis rather than a separate or opposing field. It extends the aims of traditional cybersecurity and AI safety by focusing on reasoning-level risks under adversarial pressure. All three share the objective of dependable, accountable systems, and they often reuse one another’s controls.

\paragraph{Traditional cybersecurity (CIA triad).}
Security programs implement controls including authentication, authorization, encryption, isolation, and supply-chain assurance to protect identities, data, and services. These controls remain necessary foundational elements. However, when adversaries manipulate system outputs through legitimate interfaces without violating access controls, technical security measures alone cannot ensure decision integrity. This limitation arises because the traditional CIA security objectives do not encompass the full spectrum of cognitive-level risks that emerge when AI systems process and synthesize information.

\paragraph{AI safety (alignment and reliability).}
AI safety research targets alignment, interpretability, and reduction of harmful behaviors in non-adversarial or mixed settings \cite{amodei2016concrete, christiano2018supervising}. Methods such as Constitutional AI \cite{bai2022constitutional} and RLHF \cite{ouyang2022training, christiano2017deep} materially reduce baseline errors and improve normative behavior. Cognitive cybersecurity builds on these advances: aligned models are preferable starting points, yet adversarial inputs can still exploit reasoning patterns even when average-case behavior is strong.

\paragraph{Cognitive cybersecurity.}
Our focus is adversarial exploitation of reasoning mechanisms (e.g., prompt injection, context poisoning, source interference) and its operational governance. We therefore couple CIA with Trust and Autonomy, provide a mapping to OWASP/ATLAS, and introduce CPT to test residual risk under pressure. The relationship is analogous to safety vs.\ security in classical systems, overlapping and mutually reinforcing, not mutually exclusive.

\paragraph{Layered example (diagnostic AI).}
A system analyzing medical images benefits simultaneously from (i) cybersecurity controls that protect data/system integrity and access (e.g., identity, encryption, provenance, isolation); (ii) safety alignment that improves default diagnostic behavior and calibrated uncertainty; and (iii) cognitive-security controls that protect decision/epistemic integrity by resisting adversarial prompts or context manipulations that could steer conclusions without altering inputs or bypassing access controls. The layers are coordinated: safety reduces baseline error; cybersecurity constrains who can change what and preserves content integrity; cognitive cybersecurity tests and gates deployment against adversarial steering of reasoning (e.g., prompt injection, context poisoning, source interference) so that diagnostic justifications and cited evidence remain trustworthy.

Cognitive cybersecurity complements rather than replaces existing frameworks. Organizations should layer cognitive security controls atop traditional security (authentication, encryption) and AI safety measures (RLHF, constitutional AI). The CIA+TA framework suggests a missing operational layer for reasoning-level threats.

\subsection{The Cognitive Security Suite: Seven Fundamental Vulnerabilities}

Our empirical research identified a non exhaustive list of seven categories of cognitive vulnerabilities that constitute the primary attack surface for AI systems. Each vulnerability corresponds to specific exploitation patterns observed across 12,180 controlled experiments.

Our taxonomy builds upon cognitive psychology research \cite{kahneman2011thinking, tversky1974judgment} and recent discoveries of LLM-specific vulnerabilities \cite{zou2023universal, zhou2024easyjailbreak}:

\textbf{CCS-1: Authority Hallucination} emerges when systems generate false information while maintaining authoritative presentation. Under pressure to provide comprehensive responses, systems may fabricate citations, statistics, or technical details that appear plausible but lack factual basis. This vulnerability aligns with research on hallucination in language models \cite{xu2024hallucination} and represents a critical trust challenge for AI systems. Baseline DOI validity can vary widely across models. Note: In this paper we treat validity and refusal as separate phenomena.

\textbf{CCS-2: Context Poisoning} involves the gradual corruption of reasoning through accumulated biased information across extended interactions. Adversaries introduce subtle distortions that compound over multiple exchanges, shifting the system's analytical stance without triggering anomaly detection. Measured drift was quantified as per-turn stance slopes; in released runs, means were near zero with occasional positive slopes, and we did not report standard-deviation-per-turn rates. Drift accumulates over multiple exchanges and is architecture-dependent. When retrieval is in-scope, tainted knowledge bases further compound drift through repeated exposure \cite{zou2024poisonedrag}.

\textbf{CCS-3: Goal Misalignment Loops} occur when systems cannot reconcile conflicting objectives, producing outputs that satisfy neither goal effectively. When presented with contradictory optimization targets, systems may generate inconsistent or incomplete outputs. The severity is architecture-dependent.

\textbf{CCS-4: Identity and Role Confusion} enables inappropriate persona adoption that overrides safety training and operational boundaries. Through sophisticated role-play instructions, adversaries can cause systems to violate constraints by adopting personas with different permission sets or knowledge domains. Properly implemented boundary controls achieved mitigation success rates exceeding 90\% ($\eta > 0.9$), suggesting this vulnerability can be effectively managed with appropriate measures. The effectiveness of boundary controls against identity confusion contrasts with the challenge of defending against more sophisticated prompt injection attacks \cite{liu2023prompt, liu2024automatic}.

\textbf{CCS-5: Memory and Source Interference} causes systems to incorporate false contextual claims into their reasoning, treating injected misinformation as authoritative ground truth. This vulnerability proved particularly concerning due to documented backfire effects where certain mitigation attempts amplified the vulnerability, with mitigation coefficients reaching $\eta = -1.35$ (135\% increase in error rates). In RAG pipelines, the retrieval layer constitutes an additional attack surface (not tested in our AI experiment). Prior work shows backdoored contexts can deterministically steer outputs in deployed LLMs \cite{cheng2024trojanrag}.

\textbf{CCS-6: Cognitive Load Overflow} results in degraded reasoning quality when systems face information overload that obscures critical content within verbose or irrelevant material. We define action density as the ratio of actionable items to total response tokens:

\begin{equation}
\text{ActionDensity} = \frac{\text{Number of actionable items}}{\text{Total tokens in response}}
\end{equation}

As practical diagnostics, we use heuristics such as low action density, elevated readability grade, and fact omission indicators (implementation-dependent). Mitigation effectiveness varied by architecture; Table~2 reports the aggregated median ($\tilde{\eta}=0.67$), with per-model variability captured in the released logs.

\textbf{CCS-7: Attention Hijacking} occurs when emotional framing or salience manipulation overrides analytical reasoning, producing different recommendations for logically equivalent scenarios based solely on presentation. This vulnerability demonstrated architecture sensitivity, with mitigation coefficients up to $\eta = 0.96$ (96\% reduction).

These vulnerabilities share three characteristics that distinguish them from traditional security concerns: they operate through legitimate system functions rather than technical exploits, they target reasoning processes rather than data or infrastructure, and they exhibit strong architecture-dependent patterns that prevent universal defensive strategies. This taxonomy proposes a foundation for systematic risk assessment and targeted mitigation strategies tailored to specific system architectures and deployment contexts.

\subsection{Alignment with Security Standards}

The CCS-7 taxonomy maps to established security frameworks, facilitating operational integration. As shown in Table 1, our vulnerabilities align with the OWASP Top 10 for LLMs \cite{owasp2025top10} and MITRE ATLAS \cite{mitre2021atlas}, enabling security teams to leverage existing threat intelligence while extending coverage to cognitive-specific vulnerabilities.

\begin{table*}[h]
\centering
\caption{Best-fit mapping of CCS-7 to OWASP Top 10 for LLM/GenAI (2025) and MITRE ATLAS techniques}
\small
\begin{tabular}{|p{4cm}|p{6.3cm}|p{6.3cm}|}
\hline
\textbf{CCS Category} & \textbf{OWASP LLM Top 10 (2025)} & \textbf{MITRE ATLAS (Technique IDs)} \\
\hline
CCS-1: Authority Hallucination 
& LLM09:2025 \textit{Misinformation} 
& AML.T0048.002 \textit{Societal Harm} \\
\hline
CCS-2: Context Poisoning 
& LLM04:2025 \textit{Data and Model Poisoning} 
& AML.T0020 \textit{Poison Training Data};\quad AML.T0018 \textit{Manipulate ML Model} \\
\hline
CCS-3: Goal Misalignment 
& LLM06:2025 \textit{Excessive Agency} 
& AML.T0051.000 / .001 \textit{LLM Prompt Injection (Direct / Indirect)};\quad AML.T0054 \textit{LLM Jailbreak} \\
\hline
CCS-4: Identity Confusion 
& LLM01:2025 \textit{Prompt Injection} 
& AML.T0051.000 / .001 \textit{LLM Prompt Injection (Direct / Indirect)} \\
\hline
CCS-5: Source Interference 
& LLM08:2025 \textit{Vector and Embedding Weaknesses};\newline
LLM05:2025 \textit{Improper Output Handling} 
& AML.T0020 \textit{Poison Training Data};\quad AML.T0051.001 \textit{LLM Prompt Injection: Indirect} \\
\hline
CCS-6: Cognitive Overflow 
& LLM10:2025 \textit{Unbounded Consumption} 
& AML.T0029 \textit{Denial of AI Service};\quad AML.T0034 \textit{Cost Harvesting} \\
\hline
CCS-7: Attention Hijacking 
& LLM01:2025 \textit{Prompt Injection} 
& AML.T0054 \textit{LLM Jailbreak} \\
\hline
\end{tabular}
\medskip

\begin{minipage}{0.95\textwidth}
\footnotesize
\textit{Note:} This mapping enables security teams to extend existing OWASP/MITRE-based controls with cognitive-specific mitigations. For instance, teams addressing LLM01:2025 should additionally implement CCS-4 and CCS-7 controls; teams focused on RAG/source integrity should consider CCS-5 alongside LLM08/LLM05 and AML.T0020/AML.T0051.001.
\end{minipage}
\end{table*}

\section{Quantitative Risk Assessment Framework}

\subsection{Risk Formulation}

The quantification of cognitive security risks requires a framework that captures the inherent vulnerability of AI systems and the effectiveness of implemented mitigations. We propose a two-stage risk assessment model validated through empirical measurement across 12,180 controlled experiments.

Our risk formulation adapts the NIST AI Risk Management Framework \cite{tabassi2023ai} to cognitive-specific vulnerabilities:

The \textbf{Inherent Risk} for a cognitive vulnerability $v$ is calculated as:

\begin{equation}
\text{InherentRisk}(v) = \text{norm}(E \times I \times \kappa)
\end{equation}

The normalization function employs min-max scaling to map raw scores to a 0-10 range:
\begin{equation}
\text{norm}(x) = 10 \times \frac{x - \min_{v}(E \times I \times \kappa)}{\max_{v}(E \times I \times \kappa) - \min_{v}(E \times I \times \kappa)}
\end{equation}

\paragraph{Normalization example (from Table~2 data).}
Let $E{=}0.931$, $\kappa{=}1.019$, $I{=}1$ for CCS-1 (Authority Hallucination; conditional rubric).
Then
\[
R_{\text{raw}} \;=\; 0.931\times 1.000\times 1.019 \;=\; 0.949.
\]
With $R_{\min}=0.234$ (CCS-4) and $R_{\max}=0.999$ (CCS-6), the normalized score is
\[
\text{InherentRisk} \;=\; 10 \cdot \frac{0.949-0.234}{0.999-0.234}
\;\approx\; 10 \cdot \frac{0.715}{0.766} \;\approx\; 9.34 \, .
\]

where:
\begin{itemize}
\item $E$ represents exploitability at fixed attacker budget, measured as the proportion of successful attacks under controlled conditions
\item $I$ quantifies impact on protected tasks, assessed through deviation from baseline performance metrics
\item \(\kappa\) is an architecture modifier capturing system-specific susceptibility
\item $\text{norm}$ scales the result to a 0-10 range for standardized comparison
\end{itemize}

The \textbf{Residual Risk} after applying mitigation $m$ is:

\begin{equation}
\text{ResidualRisk}(v,m) = \text{InherentRisk}(v) \times (1 - \text{ME}(m|v))
\end{equation}

where Mitigation Effectiveness (ME) is empirically derived from the change in Attack Success Rate:

\begin{equation}
ME(m|v) = 1 - \frac{ASR_{\text{attack+mit}}}{ASR_{\text{attack}}}
\end{equation}

This formulation explicitly accounts for negative values (backfire), and is consistent with Eq. (6).

This formulation also aligns with international AI risk-management guidance, ISO/IEC 23894:2023 \cite{iso23894}, which recommends explicit separation of inherent and residual risk and traceable controls.

\subsection{Architecture-Dependent Modifiers}

The architecture modifier $\kappa$ varies by design; in our AI experiment (aggregated as p75/median across models) values cluster near $1.0$ (Table~2). Architecture dependence nonetheless remains critical: a mitigation effective on one family may be counterproductive on another, for example, verification-focused prompts ranged from no measurable reduction ($\eta\approx0$) in some architectures to a 135\% increase in others ($\eta=-1.35$).

\subsection{Risk Assessment}

\begin{table*}[h]
\centering
\setlength{\tabcolsep}{4pt} 
\caption{Risk rubric computed from AI Experiment (E = median ASR under attack/no-mitigation; $\kappa$ = p75/median dispersion across models; $I{=}1$ for all CCS). Normalization to 0-10 is linear over the seven CCS raw scores observed in these logs.}
\small
\begin{tabular}{|l|c|c|c|c|c|c|}
\hline
\textbf{Vulnerability} & \textbf{$E$ (median ASR)} & \textbf{$I$} & \textbf{$\kappa$ (p75/median)} & \textbf{Inherent Risk} & \textbf{$\tilde{\eta}$ (median)} & \textbf{Residual Risk} \\
\hline
CCS-1: Authority Hallucination & 0.931 & 1.00 & 1.019 & 9.33 & $0.00$ & 9.34 \\
CCS-2: Context Poisoning        & 0.867 & 1.00 & 1.058 & 8.91 &  0.08 & 8.23 \\
CCS-3: Goal Misalignment        & 0.667 & 1.00 & 1.000 & 5.65 &  0.00 & 5.65 \\
CCS-4: Identity Confusion       & 0.217 & 1.00 & 1.077 & 0.00 &  1.00 & 0.00 \\
CCS-5: Source Interference      & 0.559 & 1.00 & 1.132 & 5.21 & $-0.32$ & 6.87 \\
CCS-6: Cognitive Overflow       & 0.933 & 1.00 & 1.071 & 10.00 & 0.67 & 3.33 \\
CCS-7: Attention Hijacking      & 0.944 & 1.00 & 1.029 & 9.64 & 0.02 & 9.41 \\
\hline
\end{tabular}

\vspace{2pt}
\begin{minipage}{0.96\textwidth}
\footnotesize
\textit{Notes.} $ASR$ is defined per vulnerability following the AI experiment scoring rubrics (e.g., invalid-DOI rate for CCS-1; planted-error adoption for CCS-5; stance-drift thresholding for CCS-2). $\tilde{\eta}$ is the median per-model mitigation effectiveness (negative values indicate backfire). Derived source data and code are released as supplemental material. For CCS-1, $E$ and $\kappa$ are computed conditional on DOI emission.
\end{minipage}
\end{table*}

Computed rubric highlights:

(1) Highest residual risks in these logs are CCS-7 (9.41) and CCS-1 (9.34), with CCS-2 close behind (8.23); 

(2) Identity Confusion (CCS-4) shows strong guardrail effectiveness in this setup (median $\tilde{\eta}=1.00 \Rightarrow$ residual risk $=0$); 

(3) Source Interference (CCS-5) exhibits backfire (median $\tilde{\eta}=-0.32$), yielding a residual risk (6.87) that exceeds its inherent risk (5.21) despite mitigation attempts.

\subsection{Empirical Validation}

The framework's validity rests on empirical testing across 12,180 controlled trials, structured to isolate and measure each component of the risk calculation.

Each vulnerability category (CCS-1 through CCS-7) underwent from 630 to 3,150 trials to ensure sufficient sample size for detecting meaningful differences across architectures. Success rates were measured through objective criteria, DOI validity for authority hallucination, stance deviation for context poisoning, decision accuracy for goal misalignment. The reported exploitability values represent the median success rate across architectures at standardized attacker effort levels.

Impact measurements evaluated the consequence of successful exploitation on representative tasks. For authority hallucination, we assessed DOI validity; for context poisoning, we used stance-shift slopes, for goal misalignment, we used decision accuracy on specified tasks. Context poisoning impact was assessed by the magnitude of stance shift under adversarial context and the per-step drift slope within turn-indexed test sequences.

Each mitigation strategy underwent controlled testing across three conditions (baseline, attack, mitigated). The mitigation coefficient $\eta$ was calculated as:

\begin{equation}
\eta = 1 - \frac{\text{ASR}_{\text{attack+mitigation}}}{\text{ASR}_{\text{attack}}}
\end{equation}

This approach to measuring mitigation effectiveness extends adversarial robustness metrics \cite{carlini2017towards} to cognitive vulnerabilities.

Negative $\eta$ values, indicating backfire effects, were observed in a non-trivial subset of architecture-mitigation combinations, concentrated primarily in CCS-5 and CCS-7. This mirrors findings from LM red-teaming studies, where naïve adversarial training can entrench or mask failure modes rather than eliminate them \cite{ganguli2022redteaming, hubinger2024sleeper}.

By incorporating architecture-specific modifiers and validated mitigation effectiveness, the framework contributes to targeted resource allocation and informed selection of defensive strategies appropriate to specific system deployments.

\section{CIA+TA Security Objectives}

The traditional CIA triad (Confidentiality, Integrity, and Availability) while foundational to cybersecurity \cite{whitten1999johnny}, requires an extension to address cognitive systems that generate knowledge claims and influence human decisions \cite{chen2024developing}. We propose the CIA+TA framework, which extends the traditional principles to incorporate cognitive risks and introduces Trust and Autonomy as security requirements.

\subsection{Recontextualized CIA for Cognitive Channels}

\textbf{Cognitive Confidentiality} extends beyond data protection to encompass the inadvertent exposure of reasoning patterns and internal knowledge structures. While traditional confidentiality prevents unauthorized data access, cognitive confidentiality addresses information leakage through inference patterns and behavioral signatures.

Crafted query sequences can reveal training and architectural characteristics (e.g., via response patterns or confidence distributions) without violating access controls. Mitigation therefore focuses on monitoring query semantics and response patterns, not only access logs.

Implementation requires differential privacy mechanisms for model outputs, providing mathematical bounds on information disclosure through repeated queries. Query pattern analysis can identify reconnaissance activities by examining semantic relationships and temporal patterns that indicate systematic probing of model boundaries.

\textbf{Cognitive Integrity} preserves accurate reasoning capabilities against adversarial corruption. Unlike data integrity, which prevents unauthorized modifications, cognitive integrity defends against legitimate inputs designed to corrupt reasoning processes.

Goal misalignment (CCS-3) produced decision deviations that compounded through feedback loops, producing amplified deviations in certain scenarios.

Preservation requires continuous validation through benchmark testing against established performance baselines, drift detection algorithms that identify gradual reasoning degradation, and cross-reference verification against authoritative sources. The challenge lies in distinguishing legitimate edge cases that expand capabilities from adversarial inputs that corrupt reasoning.

\textbf{Cognitive Availability} ensures reasoning quality remains above usable thresholds. While traditional availability focuses on uptime, cognitive availability addresses attacks that degrade output quality while leaving systems technically operational.

Cognitive load overflow (CCS-6) demonstrated this vulnerability, where information flooding reduced action density while systems continued generating responses. These degraded outputs appeared superficially reasonable but omitted critical information or contained logical inconsistencies. Such attacks prove more damaging than traditional denial-of-service because they may go undetected while corrupting downstream decisions.

Defense requires quality metrics examining factual accuracy, logical consistency, and confidence calibration. When degradation is detected, systems can implement graceful degradation protocols that prioritize critical functions or switch to simplified operation modes that maintain reliability.

\subsection{Trust: Epistemic Validation Requirements}

Trust represents justified confidence in AI-generated knowledge claims when facing adversarial manipulation. This principle addresses the challenge of validating information that appears well-reasoned but may contain subtle falsehoods or manipulated logic. The importance of trust in AI systems has been extensively studied \cite{chen2024developing}, with recent research highlighting the need for transdisciplinary approaches \cite{nature2025trust}.

The criticality of trust becomes evident through authority hallucination (CCS-1) results, where systems generated plausible but entirely fabricated citations. Without epistemic validation, falsehoods can propagate through decision chains; in our released system-level tests for CCS-5, models sometimes adopted planted errors at substantial rates, and certain guardrail prompts increased adoption (backfire), underscoring the need for independent verification.

Trust preservation requires three complementary mechanisms:

- Maintaining chains of custody for information sources throughout reasoning processes. Each knowledge claim should be traceable to its origin, whether training data, retrieval corpus, or inference from established facts.

- Providing accurate uncertainty quantification that reflects both model confidence and potential adversarial influence. Our experiments showed that properly calibrated confidence scores flagged a substantial portion of hallucinated content through anomalous certainty patterns.

- Cross-referencing generated claims against independent knowledge sources. This verification operates through separate channels from primary reasoning to prevent coordinated manipulation.

These mechanisms address hallucination modes in LLMs (spanning intrinsic, extrinsic, and retrieval-induced variants) \cite{huang2023hallucination}.

\subsection{Autonomy: Human Agency Preservation}
Autonomy ensures human decision-makers maintain independent judgment when interacting with AI systems.

This principle recognizes the fundamental challenge of preserving human agency in AI-mediated decisions, a concern that intersects with cognitive biases in decision-making \cite{kahneman1979prospect} and persuasion psychology \cite{cialdini2021influence}.

Our human-subject study demonstrated that a brief TFVA micro-lesson (3 minutes) produced statistically significant, practically meaningful improvements in security-relevant decision-making (+7.87 percentage points overall), with the largest relative gains in Ethical Responsibility (+44.4\%) and Integrity (+25.3\%), see \cite{aydin2025thinkfirstverifyalways}. These results motivate Autonomy as a major requirement. The challenge is compounded by research showing that deceptive behaviors can persist through standard safety training \cite{hubinger2024sleeper}, indicating that human-in-the-loop controls must be robust against sophisticated manipulation. Consistent with usable-security practice, we recommend interface-level cognitive-friction mechanisms (e.g., deliberate pause/confirm flows) to support reflective judgment, building on established usable security principles \cite{cranor2005security} and research on reducing AI overreliance \cite{bucinca2021trust}.

\subsection{Operational Deployment Guidelines}

Organizations interested in Trust and Autonomy can consider this operational deployment guidelines.

\paragraph{Pre-deployment Assessment Protocol}
\begin{enumerate}
\item Conduct a CCS-7 vulnerability assessment using Table~2 baseline rates to initialize priors ($E$, $\kappa$, $\tilde{\eta}$).
\item Estimate architecture-specific modifiers by targeted tests: $\kappa$ (susceptibility) and mitigation coefficient $\eta$.
\item Compute residual risks per vulnerability via Eq. Residual Risk and rank for treatment.
\item Document Trust mechanisms (provenance tracking, independent source verification, confidence calibration).
\item Implement Autonomy-preserving interfaces (deliberate pause/confirm flows, explicit source attribution).
\item Map prioritized vulnerabilities to OWASP/ATLAS controls (Table~1) and select mitigations accordingly.
\end{enumerate}

\paragraph{Risk-Based Deployment Criteria}
Based on the empirical distribution in Table 2, we propose the following operational bands (calibrate per domain):

\begin{table*}[!t]
\centering
\caption{Risk-Based Deployment Criteria}
\label{tab:deployment-criteria}
\renewcommand{\arraystretch}{1.2}
\begin{tabular}{|c|p{4.5cm}|p{7.5cm}|}
\hline
\textbf{Residual Risk*} & \textbf{Deployment Recommendation} & \textbf{Required Controls} \\
\hline
$< 5.0$ & Standard deployment & Regular monitoring; incident response on deviation \\
\hline
$5.0$--$7.0$ & Conditional deployment & Enhanced monitoring; mandatory Trust mechanisms; user warnings \\
\hline
$7.0$--$9.0$ & High-risk deployment & Continuous CPT; restricted use cases; mandatory human oversight \\
\hline
$> 9.0$ & Requires mitigation & Deploy only after additional architectural controls reduce risk \\
\hline
\multicolumn{3}{|p{14cm}|}{\textbf{Architecture flag:} if any tested mitigation yields $\eta < -0.20$, conduct architecture-specific review before go-live.}\\
\hline
\multicolumn{3}{|p{14cm}|}{\textbf{* Thresholds Definition:} Provided as illustration only.}\\
\hline
\end{tabular}
\end{table*}

\textbf{Note.} Negative $\eta$ values indicate \emph{backfire}. For CCS-5 (median $\tilde{\eta}=-0.32$), verification-focused prompts were contraindicated in most tested architectures; prefer data/knowledge-base hardening and retrieval controls over generic verification prompts.

\section{Empirical Validation}

\subsection{Experimental Methodology Summary}

The empirical foundation for the cognitive cybersecurity framework rests on two studies designed to isolate, measure, and validate cognitive vulnerabilities across diverse AI architectures.

\textbf{Study 1: Human Validation}. A human subject study with 151 participants validated the real-world impact of cognitive vulnerabilities and the effectiveness of proposed mitigations \cite{aydin2025thinkfirstverifyalways}. Participants, recruited across expertise levels, interacted with AI generated outputs under controlled manipulation conditions. The study measured susceptibility to AI-mediated influence and the protective effect of cognitive friction mechanisms.

\textbf{Study 2: Systematic Vulnerability Assessment}. We conducted 12,180 controlled experiments across seven distinct AI architectures \cite{aydin2025cognitivecybersecurityartificialintelligence}, including instruction-tuned models, base language models, and specialized reasoning systems.

The experimental design employed three conditions for each vulnerability:
\begin{itemize}
\item \textit{Baseline}: Normal operation without adversarial inputs
\item \textit{Attack}: Exposure to crafted adversarial prompts targeting specific vulnerabilities
\item \textit{Mitigated}: Attack condition with defensive interventions applied based on the ``Think First, Verify Always'' protocol
\end{itemize}

Standardized metrics enabled cross-architecture comparison: DOI validity rates for authority hallucination, stance deviation scores for context poisoning, decision accuracy for goal misalignment, and role adoption rates for identity confusion. All experiments used fixed decoding parameters held constant across runs (temperature=0.4; Tokens=500) to ensure reproducibility while maintaining realistic output diversity.

\subsection{Key Findings}

\textbf{Architecture-Dependent Vulnerability Patterns}. The most significant finding was the strong architecture dependence of both vulnerabilities and mitigation effectiveness. Identical defensive measures produced opposite effects across different architectures, invalidating universal security approaches.

For source interference (CCS-5), verification-focused prompts, intuitively expected to reduce errors, ranged from no measurable reduction ($\eta=0$) in the best case to a 135\% increase in error rates in others ($\eta = -1.35$). This backfire effect occurred in a non-trivial subset of tested architecture-mitigation combinations. This phenomenon relates to broader challenges in AI alignment, where well-intentioned interventions can produce unexpected negative consequences \cite{sharma2023towards}.

The architecture sensitivity was particularly pronounced for attention hijacking (CCS-7), where mitigation coefficients up to $\eta = 0.96$ across architectures.

These findings extend adversarial machine learning research \cite{goodfellow2015explaining, carlini2017towards} by demonstrating that cognitive vulnerabilities exhibit even greater architecture sensitivity than traditional adversarial examples.

\textbf{Temporal Dynamics of Cognitive Attacks}. Context poisoning (CCS-2) escalated across turn-indexed sequences (with slopes varying by architecture). Stance drift accumulated within these sequences, with thresholds reached sooner in more susceptible architectures. Authority hallucinations showed immediate single-turn impact, whereas context poisoning built up over sequences.

\textbf{Mitigation Effectiveness Patterns}. Successful mitigations exhibited three characteristics:
\begin{enumerate}
\item Binary prevention for boundary-based vulnerabilities (CCS-4: Identity Confusion) achieved $\eta > 0.9$ when properly implemented
\item Graduated reduction for complexity-based vulnerabilities (CCS-6: Cognitive Overflow) showed consistent but partial effectiveness (median $\tilde{\eta}=0.67$; see Table~2), with per-model variability reflected in the released logs.
\item Architecture-specific optimization was required for reasoning-based vulnerabilities (CCS-5, CCS-7) to avoid backfire effects.
\end{enumerate}

Human validation \cite{aydin2025thinkfirstverifyalways} supports the value of cognitive-friction mechanisms and shows statistically significant, practically meaningful improvements after a brief TFVA protocol micro-lesson. This aligns with broader research on cognitive biases \cite{kahneman1979prospect} and the importance of interface design in security contexts \cite{cranor2005security}.

\subsection{Limitations and Generalizability}

Several limitations bound the interpretation and application of these findings.

\textbf{Architecture Coverage}. While seven architectures provide broad representation, the rapid evolution of AI systems means new architectures may exhibit different vulnerability patterns. The tested architectures were selected for diversity but cannot encompass all possible designs, particularly future systems with novel reasoning mechanisms.

\textbf{Ecological Validity}. Laboratory conditions, while necessary for controlled measurement, may not fully capture real-world attack sophistication. Adversaries with greater resources or motivation might develop more effective exploitation techniques than those tested. Conversely, operational constraints absent in laboratory settings might provide natural defensive advantages.

\textbf{Metric Limitations}. Quantifying cognitive vulnerabilities required operational definitions that cannot capture all manifestations. For instance, measuring authority hallucination through DOI validity provides objective assessment but may miss subtler forms of confabulation. Similarly, stance drift measurements assume linear progression, potentially overlooking non-linear manipulation patterns.

\textbf{Temporal Constraints}. Experiments necessarily operated within limited timeframes, testing interactions spanning minutes to hours. Longer-term attacks operating over days or weeks might exhibit different dynamics, particularly for context poisoning and goal misalignment vulnerabilities.

\section{Conclusion}

This paper establishes cognitive cybersecurity as a complementary discipline, at the angle of technical AI cybersecurity and AI safety, addressing vulnerabilities that emerge when machines acquire reasoning capabilities. Through systematic empirical investigation, we demonstrate that these cognitive vulnerabilities operate through fundamentally different mechanisms than traditional security threats, requiring specialized frameworks for assessment and mitigation.

The CIA+TA framework extends traditional security principles to address the unique requirements of cognitive systems. While confidentiality, integrity, and availability require recontextualization for cognitive attack channels, the addition of Trust and Autonomy as co-equal principles reflects the fundamental shift from protecting data and infrastructure to securing reasoning processes and human-AI collaboration.

This framework provides a foundation for operational deployment, though further research into architectural solutions remains necessary. Organizations may consider adopting our risk assessment methodology and CIA+TA controls, with a cognitive pentest (CPT) serving as one approach to validate deployment readiness. Future research directions could include the development of architecture-aware mitigation libraries, automated CPT pipelines, and formal verification methods for cognitive security properties. 

The extent to which cognitive cybersecurity will become essential for trustworthy AI deployment remains to be determined empirically, though current evidence suggests its importance will likely increase as AI systems gain greater autonomy. Longitudinal studies and broader deployment experience will be needed to validate these frameworks' effectiveness across diverse operational contexts.

\bibliographystyle{unsrt}
\bibliography{refs}

\end{document}